\documentclass[10pt,twocolumn,superscriptaddress,sort&compress,floatfix]{article}

\usepackage{geometry}
\usepackage{flushend}

\usepackage[numbers,sort&compress]{natbib}  
\usepackage[raggedright]{titlesec} 

\usepackage{graphicx}
\usepackage{color}
\usepackage[dvipsnames]{xcolor}
\usepackage{mathrsfs}

\usepackage[font=scriptsize,justification=justified,labelfont=bf]{caption}

\definecolor{red}{rgb}{1,0,0}
\definecolor{blue}{rgb}{0,0,1}
\definecolor{dblue}{rgb}{0.5,0.5,1}
\definecolor{green}{rgb}{0.0, 0.5, 0.0}

\usepackage[normalem]{ulem}
\newcommand{\add}[1]{#1}

\begin{document}

\twocolumn[{

\begin{raggedright}
\Large \textbf{Bridging-induced microphase separation: photobleaching experiments, chromatin domains, and the need for active reactions.}
\par
\end{raggedright}

\vspace{0.4cm}

{\large C.~A. Brackley and D. Marenduzzo}

{\scriptsize SUPA, School of Physics \& Astronomy, University of Edinburgh, Peter Guthrie Tait Road, Edinburgh, EH9 3FD, UK}

\vspace{0.4cm}

\textbf{Abstract}\vspace{0.1cm}

We review the mechanism and consequences of the ``bridging-induced attraction'', a generic biophysical principle which underpins some existing models for chromosome organisation in 3-D. This attraction, which was revealed in polymer physics-inspired computer simulations, is a generic clustering tendency arising in multivalent chromatin-binding proteins, and it provides an explanation for the biogenesis of nuclear bodies and transcription factories via microphase separation. Including post-translational modification reactions involving these multivalent proteins can account for the fast dynamics of the ensuing clusters, as is observed via microscopy and photobleaching experiments. The clusters found in simulations also give rise to chromatin domains which conform well with the observation of A/B compartments in HiC experiments. 

\vspace{0.3cm}


\vspace{-0.2cm}
\begin{center}
\rule{16cm}{.03cm}
\end{center}
\vspace{-0.5cm}

\begin{quote}
Author biographies: 

\textbf{Chris A. Brackley:} \textit{Chris has been at the University of Edinburgh since 2012. His research interests are chromatin and DNA biophysics. His work involves large scale coarse-grained molecular dynamics simulations and bioinformatics.}

\textbf{Davide Marenduzzo:} \textit{Davide has been at Edinburgh since 2006. His research interests include chromosome biophysics and active matter. Davide is known for large scale coarse-grained simulations of biological and soft matter.}

\end{quote}

\vspace{-0.5cm}

\begin{center}
\rule{16cm}{.03cm}
\end{center}

\vspace{0.35cm}

}]


Key summary points:

\begin{itemize}
\item Multivalent chromatin-binding proteins have a generic tendency to aggregate: this is known as the ``bridging-induced attraction''.
\item The bridging-induced attraction provides a framework through which to understand microphase separation of nuclear proteins and formation of transcription factories in mammalian cells.
\item The bridging-induced attraction also provides a mechanisms for the formation of chromatin domains reminiscent of topologically-associating domains.
\item Photobleaching experiments showing highly dynamic nuclear clusters can be explained by nonequilibrium models which incorporate post-translational modifications or other active processes. 
\end{itemize}

\vspace{-0.8cm}
\begin{center}
\rule{5cm}{.03cm}
\end{center}
\vspace{-0.5cm}

\subsection*{Observations from HiC and microscopy experiments}\vspace{-0.3cm}

The three-dimensional (3-D) spatial organisation of chromosomes {\it in vivo} is a topic of active research in genome biology and biophysics, and has attracted the attention of both experimentalists and modellers. This is in view of the strong link between chromatin structure and gene expression~\cite{Yu2017,Cook2018}. Chromosome organisation is also known to change during cell differentiation~\cite{Dixon2012,Bonev2017}, in senescence~\cite{Chandra2015,Zirkel2017}, and in disease~\cite{Lupianez2016}. 

An important experimental technique which has allowed unprecedented progress in the field over the last few years is HiC---a high-throughout and genome-wide version of ``chromosome-conformation-capture''~\cite{LiebermanAiden2009,Dixon2012,Rao2014,Sati2017}, which uses biochemical techniques and high-throughput next-generation sequencing to build interaction, or ``contact'', maps showing which chromosome regions are close together in 3-D space. Hi-C has provided evidence that, at large scales ($>$1~Mbp), transcriptionally active regions more often interact with other active regions, and similarly inactive regions more often interact with other inactive regions. This is consistent with phase separation between euchromatin and heterochromatin (here used synonymously with active and inactive chromatin). The term ``compartments'' is often used to describe the differently interacting regions in determined by HiC, and active and inactive regions are referred to as A and B compartments respectively~\cite{LiebermanAiden2009}. 

Higher resolution contact maps showed a hierarchy of other domains at lower length-scales. Most notably, they led to the discovery of ``topologically-associating domains'' or TADs~\cite{Dixon2012,Nora2012}. There are higher-than-average levels of interaction within TADs, but reduced interactions between TADs. Within metazoans, TADs often have a well-defined pattern of histone modifications or other chromatin features, and active and inactive regions typically form separate domains~\cite{LiebermanAiden2009,Dixon2012,Rao2014,Sexton2012}. The CCCTC-binding transcription factor (CTCF) and active transcription units (binding sites for RNA polymerase II) are enriched at TAD boundaries~\cite{Rao2014,Dixon2012}. 

Microscopy studies have also been instrumental in uncovering key features of nuclear organisation. As well as showing a segregation of active and inactive chromatin consistent with compartments, they have also revealed that a number of nuclear proteins tend to form clusters, or foci. For instance, in mammalian nuclei, RNA polymerase and its associated transcription factors have been shown to localise, forming ``transcription factories'' (average size $75$ nm~\cite{Cook1999}) which are associated with active genes, and more recently a number of transcription factors and coactivators have been found to exist in liquid-like droplets~\cite{Cho2018,Sabari2018,Chong2018}. In \textit{Drosophila} \add{and embryonic stem cells of mouse and humans~\cite{Ren2008}}, polycomb group (PcG) proteins are organized into hundreds of nanoscale clusters~\cite{Wani2016}, and in mouse, HP1 localises at chromocentres which are also reminiscent of microphase separated droplets~\cite{Larson2017,Strom2017}. In the latter two cases, these structures are associated with gene inactivation. There are also many other examples of proteins which form clusters, or ``nuclear bodies''~\cite{Cook2001}, including nucleoli, Cajal and promyoleocytyc bodies, nuclear speckles etc. Some of these bodies are more complicated than others, with some having internal structure and containing several protein species and/or an RNA component. Experiments monitoring fluorescence recovery after photobleaching (FRAP) show that nuclear bodies are very dynamic, and their constituents exchange with the soluble pool in the nucleus within minutes~\cite{Handwerger2003,Hernandez2005,Kimura2002}.
     
\begin{figure*}[!ht]
\centerline{\includegraphics[width=12.75cm]{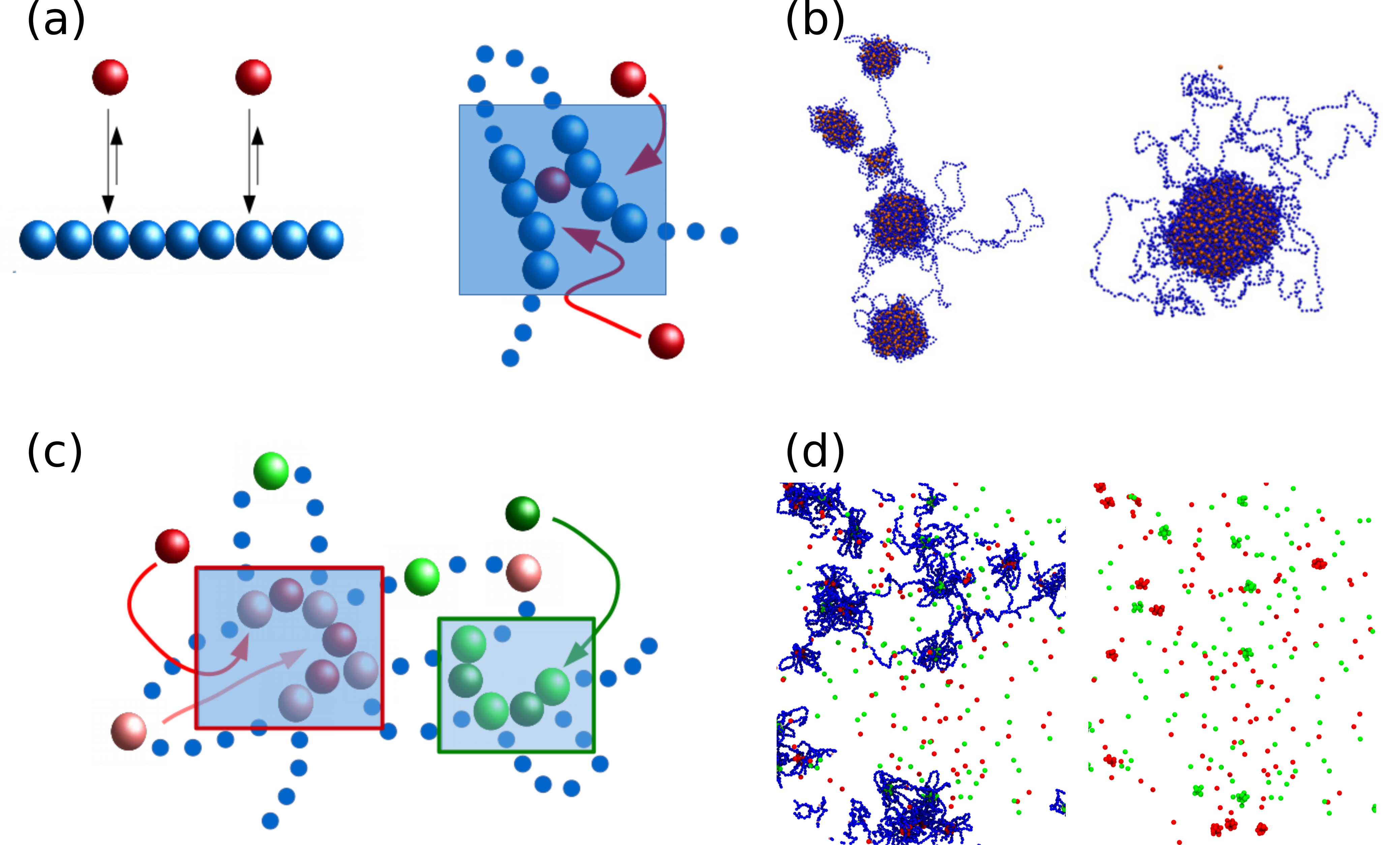}}
\caption{\label{Fig1}
\textbf{The bridging-induced attraction and (micro)phase separation. }
\textbf{(a)} A chromatin fibre is modelled as a bead-and-spring polymer, with spherical monomers (blue beads). Proteins (modelled as red spherical) bind to the chromatin fibre non-specifically. As proteins are multivalent, upon binding they create bridges which increase local chromatin density. Due to the density increase, more proteins can bind, creating a feedback loop (right image). This effect has been called the ``bridging-induced attraction''. 
\textbf{(b)} As a result of the attraction, red multivalent proteins here undergo full phase separation (two different time snapshots are shown).
\textbf{(c)} Here proteins (red spheres) interact with chromatin both non-specifically (low affinity binding to blue beads shown as dots) and specifically (high affinity binding to pink beads). Green sphere represent another kind of proteins, binding again non-specifically (blue dots) and specifically (light green beads). For both protein species the concentration of binding sites increases upon bridge formation, triggering the (separate) formation of clusters of red and green proteins via the bridging-induced attraction. 
\textbf{(d)} Now the bridging-induced attraction results in microphase separation, an example of which is shown in the left image (right image shows that same snapshot without chromatin so protein clusters are more easily seen).}
\end{figure*}

The wealth of HiC and microscopy data has stimulated the use of theoretical and computational modelling approaches, which can often help to explain the observations. There are two main ``philosophies'' which can be followed when setting up a model. One is to start from HiC data and use sophisticated fitting or machine learning procedures to generate the most likely structures which reproduce the data~\cite{Giorgetti2014,Tiana2016}---those ``inverse modelling'' approaches are reviewed, for instance, in Ref.~\cite{Serra2015}. The other philosophy is to instead start with simple biophysical principles and use these to build a ``bottom-up'', or mechanistic  model; it is this approach which we address in this review. In particular we focus on the ``transcription factor'' (TF) model, also known as the strings-and-binders model~\cite{Barbieri2012,Brackley2013,Brackley2016,Chiariello2016}. This model is based on the idea that multivalent transcription factor complexes (which can bind chromatin at more than one point to form molecular bridges) are the main genome organisers; this effects a scenario where transcription orchestrates chromosome organisation~\cite{Cook2018}. [For another example of mechanistic modelling, based on the ``loop extrusion'' principle, see the review by D. Jost.] 

\subsection*{The bridging-induced attraction and microphase separation}

The simplest version of the TF model is described schematically in Fig.~1a: a chromatin fibre (represented by a flexible bead-and-spring chain) interacts non-specifically with spheres which represent protein complexes. The latter could be complexes of transcription factors or other proteins, that can bind to two or more sites on the chromatin simultaneously, forming ``molecular bridges'' that stabilize loops. For the simple case shown in Figs.~1a-b, the TFs bind to the chromatin via a non-specific attractive interaction (i.e., they can bind at any point of the fibre). If the interaction strength is large enough \add{(equivalently, if the protein's residence time on chromatin is long enough)}, then the bound proteins spontaneously form clusters, a phenomenon first discussed in Ref.~\cite{Brackley2013}. The general principle underlying this clustering -- which occurs even in the absence of attractive chromatin-chromatin or protein-protein interactions -- was called the ``bridging-induced attraction'', as it requires multivalent binding, or bridging. 

The bridging-induced attraction works via a simple thermodynamic positive feedback loop (Fig.~1a). First, proteins bind to the chromatin; as they are multivalent, they can form bridges between different chromatin regions; this increases the local chromatin concentration. The increase in concentration facilitates further binding of proteins from the soluble pool. This results in a positive feedback (binding increases concentration which increases binding), and a cluster of proteins forms. The effect is very general, as it applies to any multivalent chromatin-binding proteins, and is appealing as a possible mechanism for the biogenesis of nuclear bodies~\cite{Brackley2017} and transcription factories~\cite{Mao2011}.

For the simple case described in Figs.~1a-b, where TFs bind non-specifically to any point along the fibre, the bridging-induced attraction generates protein clusters which continue to grow (coarsen) until there is only a single macroscopic cluster in steady state~\cite{Johnson2015} (Fig.~1b). In soft matter physics, this is known as a \textit{phase separation}~\cite{Huggins1942,Flory1942}, and here this is driven by the bridging-induced attraction. Nuclear bodies, on the other hand, only grow to a finite size: they are {\it microphase separated}, rather than fully phase separated. In the soft matter nomenclature the ``micro'' prefix implies that the phase separation process is \textit{arrested}---cluster growth is halted when they reach some finite average size determined by the system parameters.

\add{While simulations including non-specific binding proteins revealed this interesting phenomenon, these might not be relevant \textit{in vivo}. M}any transcription factors do \add{indeed} interact non-specifically with DNA and chromatin (e.g. via electrostatic interactions), \add{but this tends to be weak;} they \add{tend} have \add{much stronger} specific interactions, targeting them to cognate DNA binding sites with well-defined sequence. A more detailed TF model therefore includes stronger specific binding (e.g., in Fig.~1c red proteins bind strongly to pink chromatin beads)\add{, with non-specific interactions which would be too weak to give rise to bridge formation on their own}. With such interactions, protein clusters still form via the bridging-induced attraction mechanism since bridging increases the local concentration of both non-specific and specific binding sites. However, clusters now only grow up to a self-limiting size---they are microphase separated, and are qualitatively more similar to nuclear bodies (Fig.~1d). Arrest of cluster growth occurs because the clusters formed by the bridging-induced attraction now involve high affinity sites joined by stretches of low-affinity chromatin which tend to ``loop out'' from the cluster. Cluster growth therefore requires the creation of networks of loops---it is known from polymer physics that the entropic cost to stabilise these networks increases superlinearly with the number of high affinity sites, whereas the enthalpic gain only increases linearly~\cite{Duplantier1989,Marenduzzo2009}. 
The thermodynamic contest between this gain and cost results in microphase separation into clusters of a size where the two contributions are approximately equal~\cite{Marenduzzo2009}. Kinetically, the presence of a ``cloud of loops'' also sterically hinders the merging of two clusters, further stabilising the microphase separated structures. 

The model depicted in Figs.~1c-d includes two species of TF (shown in red and green) which have different specific binding sites. 
Intriguingly, because the bridging-induced attraction does not require any direct attraction between TFs, whenever different factors bind different chromatin segments, they form separate clusters (red and green factors can be seen in separate clusters in Fig.~1d). Here, the positive feedback loops for red and green proteins arise independently, as they involve a different set of specific binding sites. Therefore, this mechanism provides a simple explanation, for instance, for the formation of specialised factories formed by RNA PolII and PolIII (which bind different regions of the genome~\cite{Xu2008,Papantonis2013}).

\subsection*{Photobleaching experiments and protein switching}

\begin{figure*}
\centerline{\includegraphics[width=11.cm]{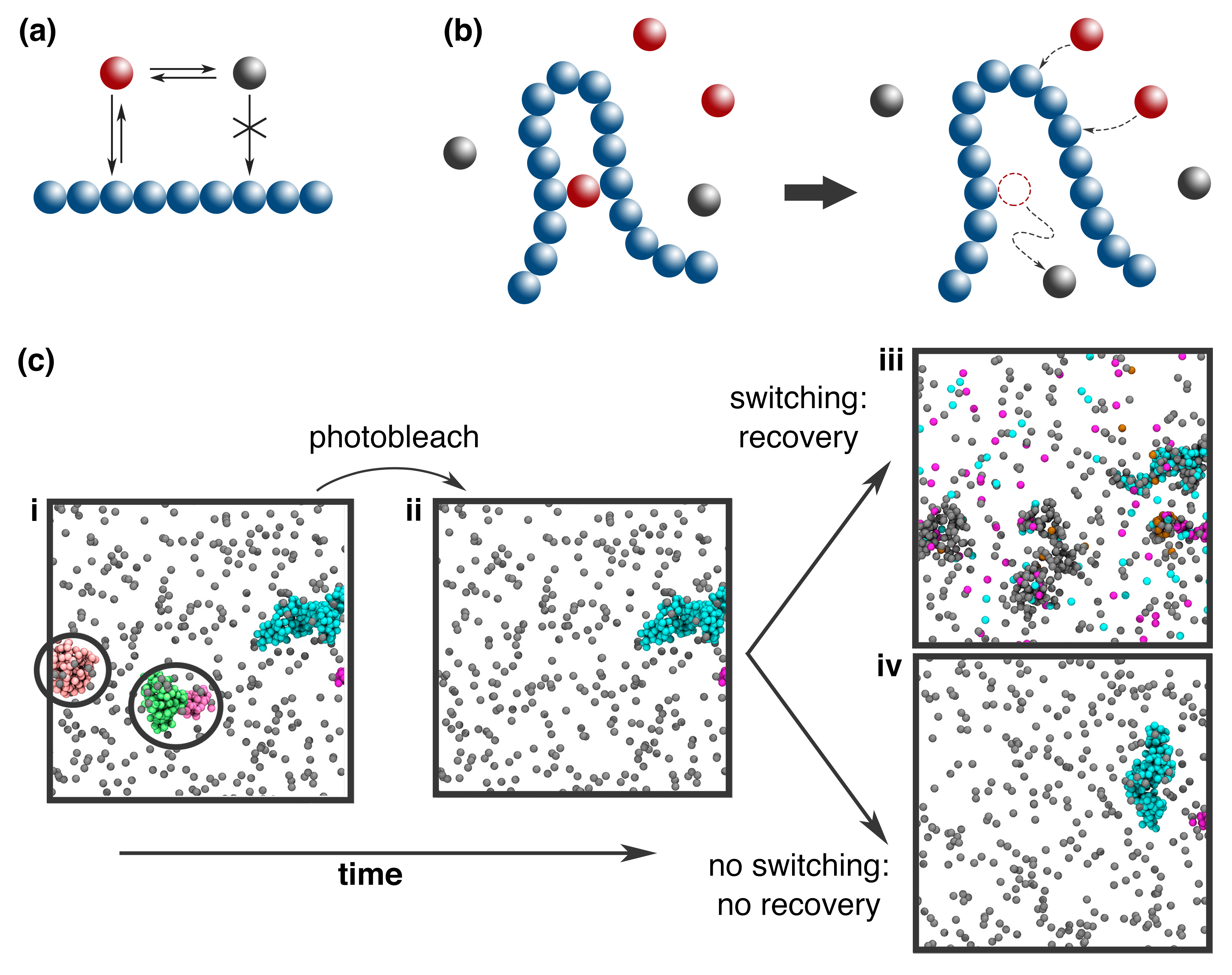}}
\caption{\label{Fig2}
\textbf{Protein switching and photobleaching in nuclear bodies.}
\textbf{(a,b)} Schematic of a TF model where proteins continually switch from an ``on'' state (chromatin-binding, red beads) to an ``off'' state (non-binding, grey beads). Switching is a generic non-equilibrium chemical reaction modelling, e.g, post-translational modification.
\textbf{(c)} Snapshots taken during a simulated FRAP experiment (for clarity only proteins, and not chromatin beads, are shown). These results were reported in Ref.~\cite{Brackley2017}. (i) At the beginning, we consider $N=2000$ non-switching proteins, half of which are able to bind the chromatin fibre (both specifically and non-specifically, see~\cite{Brackley2017} for details). The snapshot shown here is obtained after clusters have formed, and each of $5$ clusters is shown in different colours (unbound proteins are coloured grey). (ii) Photo-bleaching is simulated by making bound proteins in the highlighted circular areas invisible (although they are still included in the simulation). Hereafter colours are not changed, hence they can be used to visualise cluster dynamics. (iii) If proteins can switch, new proteins replace their ``bleached'' counterparts and clusters reappear. (iv) If proteins cannot switch, their constituent proteins do not exchange with the soluble pool and clusters do not recover. Figure adapted from~\cite{Brackley2017}, with permission.}
\end{figure*}

A further fundamental property of nuclear bodies concerns their dynamics, and is revealed by time-resolved microscopy experiments monitoring fluorescence recovery after photobleaching (FRAP). These have shown that nuclear bodies and transcription factories recover quickly (typically within minutes) after photobleaching: this means that within this time scale all of the proteins with a body have dissociated and been exchanged with proteins diffusing in the nucleoplasm. Importantly the ``turn-over'' time for the constituents of the body is orders of magnitude smaller than its lifetime. The dynamical nature of nuclear bodies found by FRAP is a challenge to explain in models: the microphase separated protein clusters found in the TF model (Fig.~1d) are static---their formation requires strong binding to chromatin, and hence the dissociation rate is very slow.

As shown in Ref.~\cite{Brackley2017}, this issue can be overcome by relaxing the assumption that the TF model works at or close to thermodynamic equilibrium. Many nuclear proteins -- including those associated with chromatin -- indeed require ATP to function, and often change their conformation in an active way to perform work. For example, RNA polymerase hydrolyses ATP as it synthesises RNA and moves along the DNA template; many other proteins interact with enzymes which add or remove chemical modifications, again in reactions involving ATP. As a simple way to model such ubiquitous post-translational modifications, we considered a scheme where proteins switch back and forth between an ``on'' (chromatin-binding) state and an ``off'' (non-binding) state. For instance, we are imagining that the wild-type of a protein binds chromatin multivalently, while its phosphorylated state does not. \add{In this scenario, the switching occurs at a rate which is independent of the binding state of the protein---when a bound protein switches off it will immediately unbind, and cannot rebind. This breaks detailed balance (and time reversibility) and drives the system away from equilibrium; in other words these are \textit{active} reactions.}

In simulations with switching proteins, clusters still form through the bridging-induced attraction. Crucially, however, the absence of thermodynamic equilibrium makes it possible to form clusters which are both stable (they form fast and reproducibly) as well as highly dynamic (they rapidly exchange their constituents with diffusing proteins in the soluble pool). This is because proteins in the ``on'' state can have arbitrarily high binding affinity to form stable clusters, and also rapidly unbind when they ``switch off''. Simulated FRAP from this model shows that clusters recover after a time close to the inverse of the switching rate, and that they retain memory of their shape for much longer time scales, just as nuclear bodies do (Fig.~2c). \add{Another effect of switching is that the probability for two regions separated by some genomic distance $s$ is reduced for large $s$, compared to the case without switching; in other words chromatin interactions become more local. This suggests that the non-equilibrium reactions lead to changes in the large scale chromatin structure as well as the dynamics of the clusters themselves.} The simple switching model is general, in that the switching could represent different post-translational modifications, active protein degradation followed by \textit{de novo} replacement, programmed unbinding of RNA polymerase at transcription termination, or any other active (i.e., non-thermodynamic) unbinding process. Since switching requires an active (non-equilibrium) reaction, a prediction of the model is that depletion of ATP should alter the dynamics of nuclear bodies and protein foci; indeed, in Ref.~\cite{Sabari2018} it was shown that ATP depletion by glucose deprivation does indeed slow down fluorescence recovery for foci of the transcriptional coactivators BRD4 and MED1. 

Finally in this section we note that \add{an equilibrium} model (without switching) could also in principle be able to account for FRAP data, but this requires fine tuning of the chromatin-protein attractive interaction to a value which is very close to, or just above, the critical attraction required for cluster formation (i.e., it must be strong enough to allow bridging, but weak enough to permit fast turnover). Whilst theoretically possible, this mechanism would therefore be much less robust than one based on active/non-equilibrium unbinding.

\subsection*{The bridging-induced attraction and chromatin domains}

In all of the simulations discussed so far, the protein clusters formed via the bridging-induced attraction are accompanied by the formation of chromatin ``domains'', in which intra-domain contacts are enriched over inter-domain contacts (e.g., in Fig.~1d one can see that the protein clusters drive the chromatin into ``clumps''). The clusters are reminiscent of TADs (and appear as such in contact maps), although this cannot be the only route through which TADs form, as many are associated with convergent CTCF loops~\cite{Rao2014}, which are not included in the current model.

To study the relationship between the bridging-induced attraction and TADs in more depth, the model in Fig.~1 can be extended to use experimental data to determine the positions of specific TF binding sites. A simple scheme to do this was proposed in Ref.~\cite{Brackley2016}: there, a whole human chromosome was simulated (Fig.~3) using a model which included two types of multivalent factor, one active 
and one inactive. 
The simulated chromatin was ``patterned'' into active-binding, inactive-binding, and non-binding regions; the active regions were identified using histone modification data (using ``chromatin states'' as mapped in Ref.~\cite{Ernst2011}), whereas inactive regions were identified simply by looking at GC content (heterochromatin and gene-poor regions largely correlate with low GC content). [In other versions of the TF model, different combinations of histone modification, protein binding or DNA-accessibility data were used to label active and inactive regions---for example, histone modification data is sufficient to predict domain and compartment patterns, but DNA-accessibility (e.g., for ATAC-seq) can better predict specific promoter-enhancer interactions at higher resolution~\cite{Brackley2016b,Buckle2018}.]

\begin{figure*}
\centerline{\includegraphics[width=11.5cm]{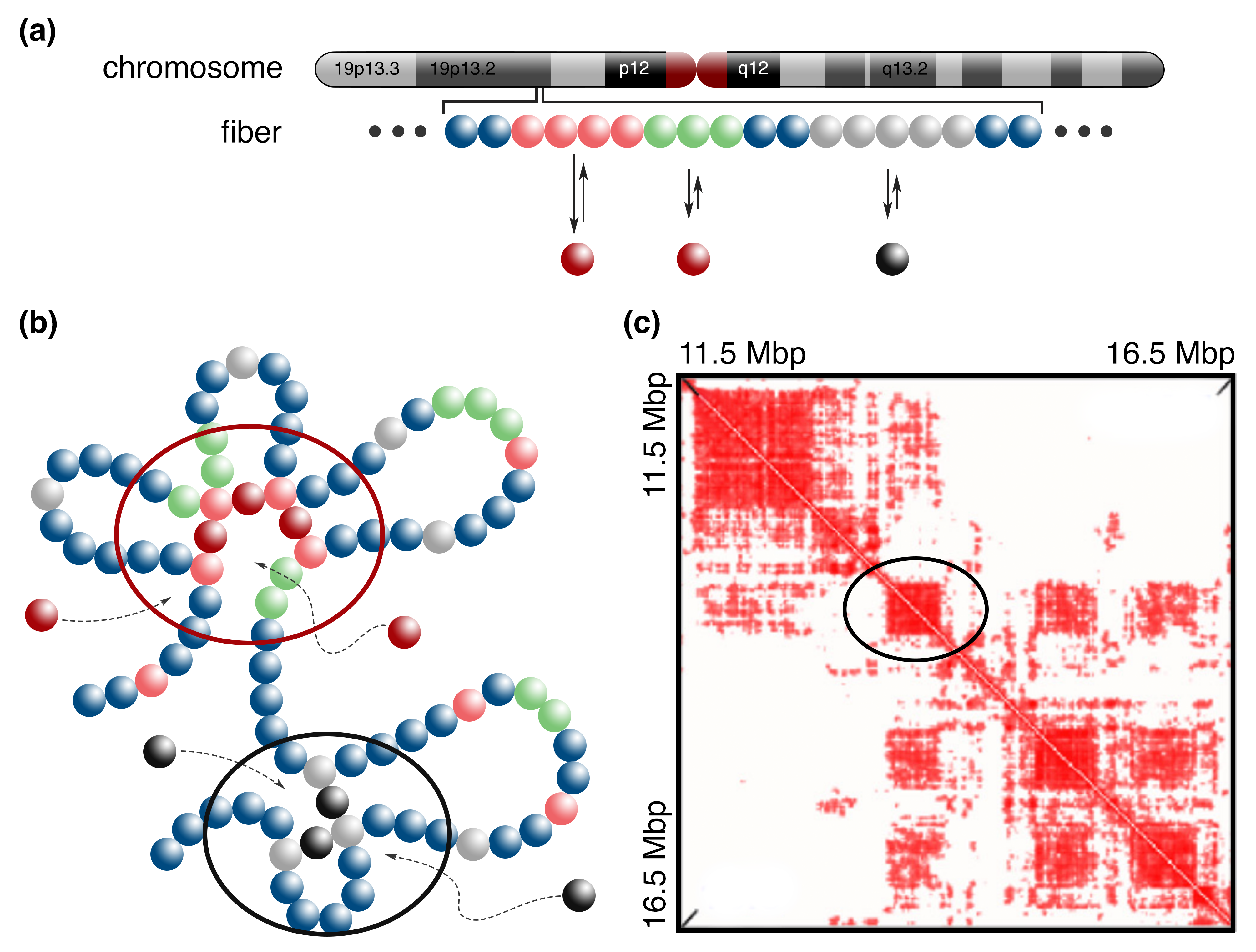}}
\caption{\label{Fig3}
\textbf{Experimental data can be used to determine binding site locations.}
\textbf{(a)} Schematic of the TF model used in Ref.~\cite{Brackley2016} to study chromosome structure. Chromatin beads are coloured according to their chromatin state as defined in Ref.~\cite{Ernst2011}, and to GC content (see~\cite{Brackley2016} for more details). Active multivalent proteins (red spheres) bind strongly (specifically) to pink beads (which are essentially promoters and enhancers~\cite{Brackley2016}) and weakly (non-specifically) to green beads (which are essentially transcribed regions). Inactive multivalent proteins (black spheres) bind weakly to grey beads (which have low GC content). Blue chromatin beads are non-binding. 
\textbf{(b)} The bridging-induced attraction drives the formation of segregated red and black protein clusters, consistent with the microphase separation of heterochromatin and euchromatin. 
\textbf{(c)} Protein clusters lead to domains in the contact map. The zoom is from a region of human chromosome 19~\cite{Brackley2016}. Adapted from~\cite{Brackley2016}, with permission.
}
\end{figure*}

Simulations performed within this model (see Fig.~3) naturally lead to microphase separation between heterochromatin and euchromatin, and the formation of structures compatible with the A/B compartments observed in HiC (Fig.~3c). Again, as in Fig.~1d, segregation between clusters occurs because different types of factor bind different chromatin regions. As well as compartmentalisation, the TF model also gives a good prediction of the locations of TAD boundaries: for example, in Ref.~\cite{Brackley2016} about 85\% of the boundaries in chromosome 19 were correctly predicted to within $100$~kbp. Also some inter-domain interactions are correctly captured (as the off-diagonal blocks in the contact maps in Fig.~3c match those in HiC maps, see Ref.~\cite{Brackley2016}). We stress that this model is predictive and fitting-free, as its only input is the 1-D information on the protein binding landscape (i.e., no HiC data is used in the model). 

We close this section by mentioning another popular model for chromosome organisation which is of a similar spirit to the TF model, namely the  ``block-copolymer'' model~\cite{Jost2014}. There, the chromatin beads interact attractively with each other directly, so bridging proteins are implied but not explicitly modelled (the attraction could represent a `bridge' protein which is tightly bound to the chromatin, or could be mediated, e.g., though interactions between histone tails and nucleosome surface charges). This approach is equivalent to the TF model if bridging proteins are abundant enough to saturate all of the binding sites; however, the two models differ in the regime where only some of the binding sites are occupied. 

\subsection*{Conclusions}

Phase separation has recently risen to become a very simple and powerful concept in chromatin and cell biology~\cite{Hyman2014,Boeynaems2018}. \add{Experiments with purified proteins have revealed, for example, that HP1 and polycomb group proteins can undergo liquid-liquid phase separation \textit{in vitro}, driven by inter-protein interactions~\cite{Tatavosian2019,Plys2019}.} Here we have reviewed a minimal model for chromatin and its associated binding proteins, \add{which revealed an additional mechanism through which phase separation can occur.} Multivalency in chromatin-binding interactions provides a generic pathway to intranuclear phase separation, even in the absence of any pre-existing interprotein interaction (Fig.~1). This ``bridging-induced attraction'' arises as multivalent binding increases local chromatin density, which in turn further increases binding, creating a generic positive feedback loop which favours clustering. 

When both specific and non-specific chromatin-protein interactions are included, the bridging-induced attraction leads to a microphase separation (arrested phase separation), and provides a simple framework within which to explain the biogenesis of nuclear bodies, transcription factories, and other protein foci. In this context, an important feature of these structures is their highly dynamic nature, which is naturally explained only by a model where proteins ``switch'' back and forth between an ``on'' state (which binds chromatin) and an ``off'' state (which is non-binding). Such switching is a simple way to simulate post-translational modification (acetylation, phosphorylation, etc.) which are routinely encountered in proteins within living cells. The action of such ``out-of-equilibrium'' chemical reactions on phase separation~\cite{Weber2019} has also attracted attention in the context of cytoplasmic structures~\cite{Brangwynne2009,Zwicker2014,Jain2016}, and is thought to play a role in protein-aggregation and misfolding based disease~\cite{Shin2017}.

The bridging-induced attraction can also provide a mechanism for formation of specialised factories of RNA PolII and PolIII~\cite{Xu2008,Papantonis2013}, and to create specificity in cluster formation more generally (different proteins form different clusters). This phenomenon is a simple consequence of this specificity in chromatin-protein interactions; it is more difficult to achieve with other phase separation models where proteins have non-specific interactions with each other. The attraction also leads to the formation of TAD-like chromatin domains which are compatible with HiC contact maps (Fig.~3). As noted above, another popular model for domain formation -- loop extrusion -- is reviewed by D.~Jost. We note that TF and extrusion are not necessarily competing models: they each explain different aspects of 3-D chromatin structure. The bridging induced attraction mechanism naturally explains chromatin compartments and the related domains, as well as protein clustering, microphase separation and dynamics; whereas loop extrusion naturally explains CTCF mediated loops, loop domains, and experiments on the function of cohesin during interphase and condensin during mitosis~\cite{Nishiyama2019,Takahashi2019}. Indeed there have been a number of studies which have combined aspects of the two models~\cite{Buckle2018,Pereira2018,Nuebler2018}.

There are several open problems in this area which it would be good to address in the near future. First, it would be of interest to study the dynamics of the proteins in the microphase separated clusters formed via the bridging-induced attraction. Presumably, different parameters may yield either liquid-like or glassy/solid-like dynamics, and it would be important to relate this to recent experimental studies~\cite{Larson2017,Strom2017}. Second, it would be desirable to make the model more realistic by including protein-protein as well as chromatin-protein interactions, and by making switching rates dependent on local 3-D structure (e.g. to model an inhomogeneous spatial distribution of the enzymes involved in the post-translational modification). \add{For example, the bridging induced attraction might act to nucleate and localise protein foci which then grow further via a interprotein interaction mediated liquid-liquid phase separation. Different protein-protein and protein chromatin interactions might lead to heterogenetiy within foci, and this might in turn lead to spatial variation in active reactions.} Finally, it would be intriguing to devise synthetic analogues of the bridging-induced attraction with DNA-protein systems {\it in vitro}, which would allow, among others, a quantitative test of the growth laws predicted by the theory~\cite{Brackley2017}.

\noindent\textbf{\textit{Acknowledgments}}
This work was supported by ERC (CoG 648050, THREEDCELLPHYSICS).
We acknowledge P. R. Cook, J. Johnson, D. Michieletto, M. C. F. Pereira, for useful discussions and for their collaboration on several of the topics reviewed here.

{\scriptsize
\bibliography{references}
}


\end{document}